\documentclass[aps,prl,onecolumn,superscriptaddress,groupedaddress]{revtex4}  
\usepackage{graphicx}  
\usepackage{dcolumn}   
\usepackage{bm}        
\usepackage{amssymb}
\usepackage{amsmath} 
\usepackage[caption=false]{subfig}
\usepackage{float} 
\usepackage{lipsum} 
\usepackage{color}
\usepackage{subfig}
\usepackage{qcircuit}
\usepackage{setspace}

\usepackage{array}

\newcolumntype{L}[1]{>{\raggedright\arraybackslash}m{#1}}
\newcolumntype{C}[1]{>{\centering\arraybackslash}m{#1}}
\newcolumntype{R}[1]{>{\raggedleft\arraybackslash}m{#1}}

\newcommand{\be}{\begin{equation}}
\newcommand{\ee}{\end{equation}}
\newcommand{\ba}{\begin{eqnarray}}
\newcommand{\ea}{\end{eqnarray}}

\def\ben{\begin{equation}}
\def\een{\end{equation}}

\def\bea{\begin{eqnarray}}
\def\eea{\end{eqnarray}}

\begin{document}
\newcount\hour \newcount\minute
\hour=\time  \divide \hour by 60
\minute=\time
\loop \ifnum \minute > 59 \advance \minute by -60 \repeat
\def\nowtwelve{\ifnum \hour<13 \number\hour:
                      \ifnum \minute<10 0\fi
                      \number\minute
                      \ifnum \hour<12 \ A.M.\else \ P.M.\fi
	 \else \advance \hour by -12 \number\hour:
                      \ifnum \minute<10 0\fi
                      \number\minute \ P.M.\fi}
\def\nowtwentyfour{\ifnum \hour<10 0\fi
		\number\hour:
         	\ifnum \minute<10 0\fi
         	\number\minute}
\def \now {\nowtwelve}

\title{MAX INDEPENDENT SET AND THE QUANTUM ALTERNATING OPERATOR ANSATZ}

\author{Zain H. Saleem}
\email{zsaleem@anl.gov}
\affiliation{Argonne National Laboratory, 9700 S. Cass Ave.,
Lemont, IL 60439, USA}

\begin{abstract}

The maximum independent set (MIS) problem of graph theory using the quantum alternating operator ansatz is studied. We perform simulations on the Rigetti Forest simulator for the square ring, $K_{2,3}$, and $K_{3,3}$ graphs and analyze the dependence of the algorithm on the depth of the circuit and initial states. The probability distribution of observation of the feasible states representing  maximum independent sets is observed to be asymmetric for the MIS problem, which is unlike the Max-Cut problem where the probability distribution of feasible states is symmetric. For asymmetric graphs it is shown that the algorithm clearly favors the independent set with the larger number of elements even for finite circuit depth. We also  compare the approximation ratios for the algorithm when we choose different initial states for the square ring graph and show that it is dependent on the choice of the initial state. 
\end{abstract}

\preprint{}

\maketitle

\section{Introduction}
The quantum computation community has been expressing  growing interest in developing algorithms that can be implemented on near-term quantum machines \cite{preskil}. Several hybrid classical-quantum algorithms \cite{farhi2014,perruzo, moll} have been proposed that can take advantage of the available quantum resources in the presence of noisy gates and small decoherence times. The Quantum Approximate
Optimization Algorithm (QAOA) \cite{farhi2014} and the Variational
Quantum Eigensolver (VQE) \cite{perruzo} are two such classical-quantum algorithms. QAOA  has been put forward to tackle combinatorial optimization problems, and the VQE algorithm has application in quantum chemistry problems where the ground state of a wave function needs to be determined. The VQE algorithm is used as a subroutine in  QAOA.

In most of the hybrid algorithms the quantum part of the algorithm involves preparing a quantum circuit, and the classical part involves optimization. In the Quantum Approximate Optimization Algorithm a quantum state is created by a p-depth circuit specified by 2p variational parameters. The algorithm has been shown to be not efficiently simulatable classically even at the lowest p=1 depth \cite{farhi2016}. QAOA is thus a good candidate algorithm to study quantum advantage on near-term quantum machines. Although one can theoretically prove the success of QAOA in the  $p\to \infty$ limit as it 
approximates adiabatic quantum annealing \cite{farhi2014} in that limit, little is known about its performance when $1 < p \ll \infty$. 

A significant amount of work on QAOA has been done in the context of the Max-Cut problem, which is an unconstrained optimization problem. However, not much work has been done on constrained combinatorial optimization problems in the quantum algorithms context \cite{shengtao}. The maximum independent set (MIS) problem is considered an ``unconstrained optimization'' problem. Unlike the Max-Cut problem, in which all the $2^n$ states are feasible, the feasible states in the MIS problem consist of a subset of the configuration space. For such ``constrained optimization'' problems a quantum alternating operator ansatz \cite{stuart2017,shengtao} has been proposed. In this paper we present a simulation of the quantum alternating operator ansatz on the Rigetti Forest simulator \cite{riggeti}.

\section{Quantum Approximate Optimization Algorithm}
The QAOA algorithm was proposed for unconstrained discrete optimization problems, such as  Max-Sat, Max-Cut, and Max-Clique. Formally, consider
\begin{align}
C(\mathbf{x}) = \sum_{i=1}^{n} C_i(\mathbf{x}),
\end{align}
where $\mathbf{x} = [x_1, x_2, \ldots, x_n]$ denotes a binary label and $C_i(\mathbf{x})$ is the $i$th binary clause. The goal in optimization problems is to find a binary vector $\mathbf{x}^*$ that maximizes the number of satisfied clauses $C_i(\mathbf{x})$.

For  unconstrained combinatorial optimization problems the quantum state is typically initialized to the superposition state $|+\rangle^{\otimes n}$. For the cost Hamiltonian $C$, let $U(C, \gamma)$ denote a unitary operator with an angle $0 \leq \gamma \leq 2\pi$, defined by 
\begin{align}
U(C, \gamma) = \exp(- i  \gamma C) = \prod^{n}_{i=1} \mathrm{e}^{- \gamma C_i}.
\end{align}
We also define a driver Hamiltonian $B=\displaystyle{\sum^n_{j=1}} X_j$, which flips $n$ qubits independently. The unitary operator for the Hamiltonian with an angle $0 \leq \beta \leq \pi$ is defined as
\begin{align}
U(B, \beta) = \exp(- i \beta B) = \prod^{n}_{j=1} \mathrm{e}^{- i \beta X_j}.
\end{align}
The ground state of the driver Hamiltonian is $|\phi\rangle = |+\rangle^{\otimes n}$. The quantum approximate optimization algorithm uses an alternating quantum circuit of depth $p$ depending on Hamiltonians $B$ and $C$ to maximize the expected cost function, with $2p$ angle parameters $\boldsymbol{\gamma}$ and $\boldsymbol{\beta}$:
\begin{align}
|{\boldsymbol{\gamma}, \boldsymbol{\beta}}\rangle =
U(B, \beta_p)U(C, \gamma_p) \cdots U(B, \beta_1)U(C, \gamma_1) |\phi\rangle.
\end{align}
If we denote expectation of the cost function $C$ as $F_p$,
\begin{align}
F_p({\boldsymbol{\gamma}, \boldsymbol{\beta}}) =
\langle C \rangle(\boldsymbol{\gamma}, \boldsymbol{\beta})
=\langle {\boldsymbol{\gamma}, \boldsymbol{\beta}} | C | {\boldsymbol{\gamma}, \boldsymbol{\beta}} \rangle,
\end{align}
and let $F^\star_p$ be the maximum of $F_p({\boldsymbol{\gamma}, \boldsymbol{\beta}})$ over the angles,
$F^\star_p = \max_{{\boldsymbol{\gamma}, \boldsymbol{\beta}}} F_p({\boldsymbol{\gamma}, \boldsymbol{\beta}})$, the objective of QAOA algorithm is to maximize $F^\star_p$ by properly choosing parameters $\boldsymbol{\gamma}, \boldsymbol{\beta}$.
The approximation improves as we increase  $p$, and at infinite depth we have $\lim_{p \rightarrow \infty} F^\star_p = \max_\mathbf{x} C(\mathbf{x})$. The  expectation $F_p({\boldsymbol{\gamma}, \boldsymbol{\beta}})$ is calculated by repeated measurements on quantum computers. The variational parameters are optimized on classical computers, for example, by using the Nelder--Mead method as part of the VQE subroutine. 

\section{Max-Cut}
The  Max-Cut combinatorial optimization problem is stated as follows: Given a graph $G=(V,E)$  with nodes $V$ and edges $E$, find a subset  $S \in V$ such that the number of edges between $S$ and $S-V$ is maximized. Finding an exact solution for the Max-Cut problem is NP-hard~\cite{karp1972a}, but efficient polynomial-time classical algorithms do exist that find an approximate answer within some fixed factor of the optimum solution.

To apply the QAOA algorithm on the Max-Cut problem, we first encode the graph of the particular problem instance into a cost Hamiltonian for which any
bit string gives an energy that is the negative of the number of the cut edges. Such a cost Hamiltonian is given by 

\begin{equation}
    C = \frac{1}{2} \sum_{i,j \in  E}w_{ij} ( 1- Z_i Z_j).
\end{equation}

Here $\sigma^z_i$ is the Pauli Z matrix applied to qubit $i$, $E$ is the set of edges, and  $w$ is the adjacency matrix of the graph, with $w_{ij}=1$  if nodes are connected and zero otherwise. Since this is an unconstrained optimization problem, the initial state is prepared as a uniform superposition of all the bit strings. The mixing Hamiltonian $B$ is just a sum of the Pauli $X_i$ matrices acting on the $i$th qubit. 

\begin{equation}
B= \sum_{i \in V}X_i
\end{equation}
\subsection{Simulation of Max-Cut QAOA}

We  simulate the QAOA algorithm on the Rigetti Forest simulator \cite{riggeti}. The simulations are performed without including the noisiness of the gates. The variational quantum eigensolver subroutine is used to find the optimized parameters $\beta$ and $\gamma$. Within the VQE we use the classical Nelder--Mead method. The algorithm is run over 50 iterations, and the arithmetic averages of the probabilities of the states over these 50 iterations is calculated. 

We choose the square ring, $K_{2,3}$, and $K_{3,3}$ graphs given in Figures \ref{image1}--\ref{image3} for our simulations. For the square ring graph the Max-Cuts are the $(1,3)$ and $(2,4)$ sets corresponding to the $\langle0101\rangle$ and $\langle1010\rangle$ states, respectively. For the $K_{2,3}$ graph, the Max-Cuts are $(1,2)$ and $(3,4,5)$ corresponding to the $\langle00011\rangle$ and $\langle11100\rangle$ states, respectively. Similarly for the $K_{3,3}$ states the Max-Cuts are the $(1,2,3)$ and $(3,4,5)$ sets corresponding to the $\langle000111\rangle$ and $\langle111000\rangle$ states, respectively. Since this is an unconstrained optimization problem, every set is a cut and represents a feasible solution. However, we expect to see the peaks in the probability distribution at the states representing the Max-Cuts.

\begin{figure}[h]
\minipage{0.18\textwidth}
  \includegraphics[width=\linewidth]{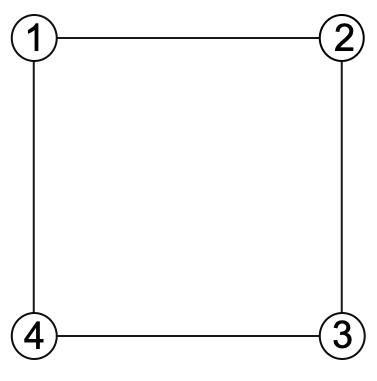}
    \caption{Square ring graph}\label{image1}

\endminipage\hfill
\minipage{0.30\textwidth}
  \includegraphics[width=\linewidth]{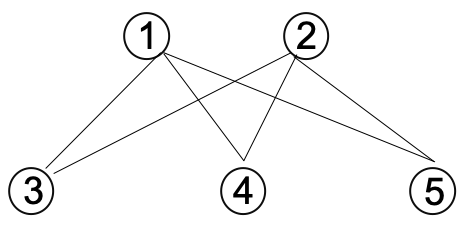}
   \caption{$K_{2,3}$ graph}\label{image2}

\endminipage\hfill
\minipage{0.30\textwidth}%
  \includegraphics[width=\linewidth]{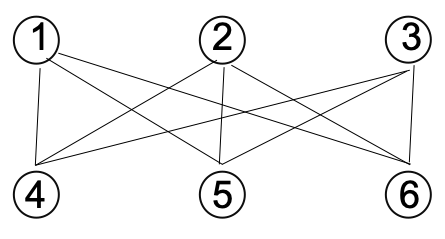}
    \caption{$K_{3,3}$ graph}\label{image3}

\endminipage

\end{figure}

The results of the simulation for the three graphs (square ring, $K_{2,3}$, and $K_{3,3}$) are provided in Figure $\ref{maxcut1}$. One can see that the peaks are located at the Max-Cut solutions. For small values of $p$, other solutions also contribute; but as the value of $p$ is increased, the Max-Cut solutions dominate, and all other peaks disappear from the distribution. We also note that the probability distribution is symmetric in the Max-Cut and the feasible solutions. 
\begin{figure}[h] 
\minipage{0.33\textwidth}
  \includegraphics[width=\linewidth]{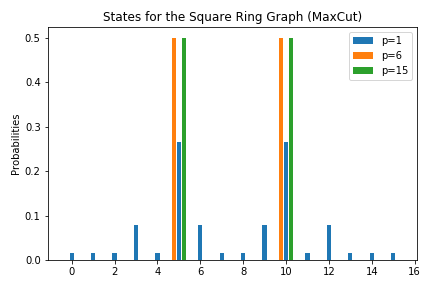}
\endminipage\hfill
\minipage{0.33\textwidth}
  \includegraphics[width=\linewidth]{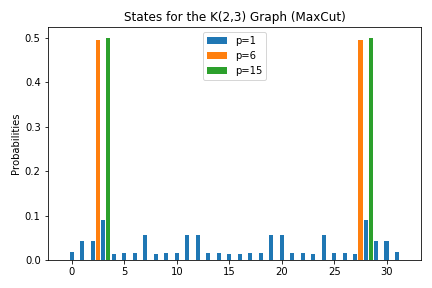}
\endminipage\hfill
\minipage{0.33\textwidth}%
  \includegraphics[width=\linewidth]{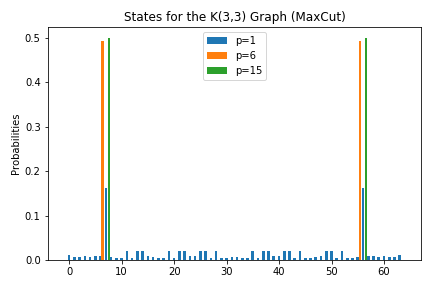}
\endminipage
\caption{Probability distribution of states for Max-Cut QAOA for the square ring, $K_{2,3}$, and $K_{3,3}$ graphs when $p=1,6$ and $15$. }\label{maxcut1}
\end{figure}

\section{Quantum Alternating Operator Ansatz}
A general QAOA circuit is defined by two parameterized families of operators: a family of phase separation operators $U_C(\gamma)$ that depends on the cost function and a family of $U_B(\beta)$ that depends on the domain and its structure. In the earlier implementation of unconstrained QAOA the feasible set of states consisted of  the entire configuration space, and therefore the mixing operator in the algorithm was $U_B(\beta)= \exp(- i \beta B)$. The constrained optimization problems, however, require optimization over feasible solutions that are typically a subset of a configuration space. The feasible solution set is specified by a set of Boolean functions (hard constraints) that are satisfied by the feasible solutions. If the mixing operators preserve feasibility, then given a feasible initial state, the QAOA algorithm will produce a final state that, when measured, gives a feasible solution. This is achieved by the quantum alternating operator ansatz, which comprises three main components: the initial state, the phase operators, and the mixing operators.
The initial state must be feasible, and it must be trivial to implement such that it can be created by a constant depth quantum circuit from the $|0...0 \rangle_n$ state. The family of mixing unitaries $U_B(\beta)$ are required to take feasible states to feasible states for all values of parameters and must also provide transitions between all feasible solutions. For an objective function $C$ we define
$H_C$ to be the Hamiltonian that acts as $C$ on basis states $H_C |\mathbf{x}\rangle = C(\mathbf{x}) |\mathbf{x}\rangle.$ The phase separation operators~$U_C(\gamma)$ are required to be diagonal in the computational basis, and therefore the phase separation unitary is defined as $U_C(\gamma) = e^{-i \gamma H_C}$ up to trivial global phase terms.
\section{Maximum Independent Set}

Consider a graph $G = (V,E)$, with $V$ the set of nodes of the graph and $E$ the set of edges. Let $\mathcal{N}(i)=\{j \in V: (i,j) \in E\}$ be the neighbors of the $i^{th}$ node in $V$. Positive weights $w_i$ are associated with each node $i$. A subset $V'$ of $V$ is represented by a vector $\textbf{x} = (x_i) \in
\{0,1\}^{|V|}$, where $x_i = 1$ means $i$ is in the subset and $x_i = 0$ means $i$ is not in the subset. A subset $\textbf{x}$ is called an {independent set} if no two nodes in the subset are connected by an edge: $(x_i, x_j) \neq (1,1)$ for all $(i,j) \in E$. The maximum independent set is the independent set with the largest number of nodes. We are interested in finding a maximum weighted independent set  $\textbf{x}^*$. 

No known polynomial-time classical algorithm  solves the maximum independent set unless P=NP \cite{Tre}. The best algorithm known for general graphs give approximations within a polynomial factor. MIS can be approximated to $(D_g + 2)/3$ \cite{a2} on bounded-degree graphs with maximum degree $D_g \geq 3$,  but it still remains APX-complete \cite{a3}. The best-known classical algorithm for the weighted maximum independent set is the greedy local search algorithm \cite{chandra}, which also gives a polynomial factor approximation. 

The three QAOA components for this maximum independent set problem are as follows.

\begin{itemize}
	\item Initial state: The initial state can be the trivial state or any state representing the independent set.
	\item Phase separation Hamiltonian: The objective function $H_C(x) = \sum_{j=1}^n x_j $ counts the number of vertices in $V'$, and the Hamiltonian corresponding to the function is 
	\begin{equation}
	H_C = \frac12  \sum_{u\in V} (I-Z_{u}). 
	\end{equation}
		
	\item Mixing Hamiltonian: When constructing the mixing Hamiltonian, we note two points: (1) given an independent set $V'$, adding a vertex $w \notin V'$ to $V'$ preserves feasibility only if none of the neighbors of $w'$ are already in $V'$; and (2) we can always remove any vertex $w \in V'$ without affecting the feasibility of the state. The transformation rule that preserves the feasibility is to flip the bit $x_w$ if and only if $\bar{x}_{v_1  } \bar{x}_{v_2 }\dots \bar{x}_{v_\ell }=1$, where $v_1,\dots,v_\ell$ are the vertices adjacent to $w$.  Keeping these observations in mind, we can construct the following Hamiltonian: $B=\sum_u B_u$, where   
	\begin{equation}  \label{eqn:driverIndepSet}
	B_{u} = \frac{1}{2^{\ell}} X_u \; 
	\prod_{j=1}^{\ell} (I+Z_{v_j }). 
	\end{equation}
	This is the Hamiltonian-based implementation of the mixing unitaries. A sequential implementation of the mixing unitaries is provided in \cite{stuart2017} and has some advantages, but we will leave that implementation for later work.
\end{itemize}

\subsection{Simulation of Maximum Independent Set}
The domain in the MIS problem is the n-bit strings corresponding to the independent sets in $G$. We again simulate the quantum alternating operator ansatz  for the square ring, $K_{2,3}$, and $K_{3,3}$ graphs. In the case of the MIS problem, not all sets are feasible solutions of the problem. For example, in the square ring graph, the independent sets are $(\phi), (1), (2), (3),(4), (1,3)$, and $(2,4)$ corresponding to the states $|0000\rangle, |0001\rangle, |0010\rangle, |0100\rangle, |1000\rangle, |0101\rangle$, and $|1010\rangle$, respectively. All other sets are not  feasible solutions. Two maximum independent sets in the square ring graph correspond to $(1,3)$ and $(2,4)$.  For the $K_{2,3}$ graph, the maximum independent sets are $(1,2)$ and $(3,4,5)$ corresponding to the $\langle00011\rangle$ and $\langle11100\rangle$ states, respectively. Similarly for the $K_{3,3}$ states the maximum independent sets are the $(1,2,3)$ and $(3,4,5)$ sets corresponding to the $\langle000111\rangle$ and $\langle111000\rangle$ states, respectively.

Below we  present the results of our simulations. The MIS problem differs from the MAX-Cut problem in two crucial ways.  (1) The probability distribution of the states is asymmetric. This is  due to the asymmetry in the mixing operator. Unlike the Max-Cut problem where the mixing operator acts symmetrically on all qubits, in the MIS problem the mixing operator acts asymmetrically. (2) The initial state in the MIS problem can be any of the independent sets. In the Max-Cut problem the choice of the initial state was obvious, whereas in the independent set problem we can choose any of the independent sets as the initial state. 

\textbf{Asymmetric probability distribution}: We analyze the probability distribution for the three graphs: square ring, $K_{2,3}$, and $K_{3,3}$. The initial state we  use is the same empty set state $\langle 00\cdots \rangle$ for the three graphs. As expected, the probability distributions shown in Figure \ref{MIS} are asymmetric in all  three cases. As the value of $p$ is increased, however, the distributions become more symmetric. The reason  is that increasing $p$ allows more mixing to take place between the feasible solutions.

\begin{figure}[ht]
\minipage{0.33\textwidth}
  \includegraphics[width=\linewidth]{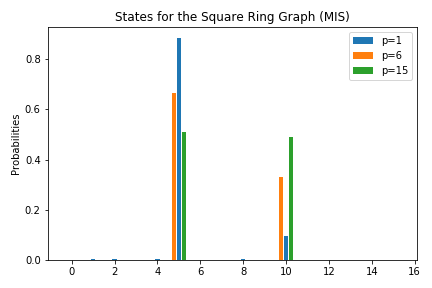}
\endminipage\hfill
\minipage{0.33\textwidth}
  \includegraphics[width=\linewidth]{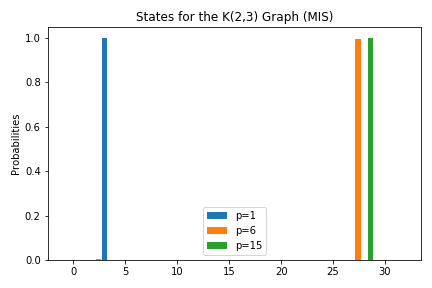}
\endminipage\hfill
\minipage{0.33\textwidth}%
  \includegraphics[width=\linewidth]{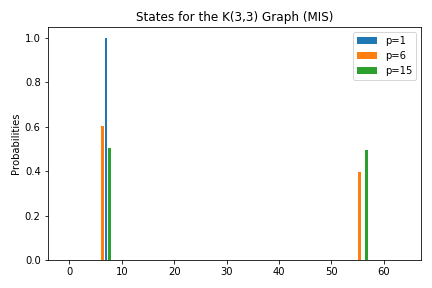}
\endminipage
\caption{Probability distribution of states for Max-Independent set QAOA for the square ring, $K_{2,3}$, and $K_{3,3}$ graphs when $p=1,6$ and $15$.}\label{MIS}
\end{figure}
The square ring and $K_{3,3}$ graphs are symmetric whereas the $K_{2,3}$ graph is an asymmetric graph. We note that for the $K_{2,3}$ graph, even when $p=6$ (finite circuit depth), the contribution from the independent set containing a larger number of elements $(3,4,5)$ is considerably larger than from any other set.

\textbf{Dependence on initial states}:
We also check the dependence of the outcome of our quantum approximate optimization algorithm  on the choice of initial states. Here we  analyze only the square ring graph. In the experiments that tested the dependence of the algorithm on the circuit depth and asymmetry of the probability distribution, we used the zero state (empty set) as our initial state. Here we  run our simulations with the $\langle 0101 \rangle$ and $\langle 1010 \rangle$ initial states. 

\begin{figure}[ht]
\centering
\minipage{0.33\textwidth}
  \includegraphics[width=\linewidth]{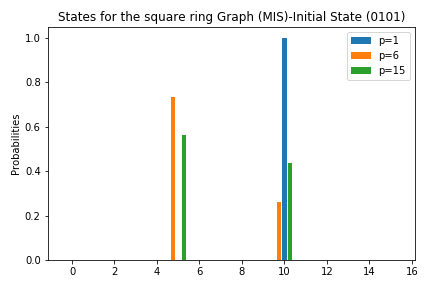}
\endminipage
\minipage{0.33\textwidth}
  \includegraphics[width=\linewidth]{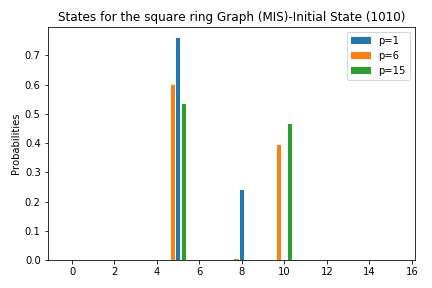}
\endminipage\hfill
\caption{Dependence of the algorithm on the initial states.}
\end{figure}
We can see that for lower values of $p$ the initial state dominates the probability distribution. As the value of $p$ is increased, however, the distribution becomes more and more symmetrical.

\section{Initial States and Approximation Ratio }
The analytical calculation of $\langle C \rangle=\langle {\boldsymbol{\gamma}, \boldsymbol{\beta}} | C | {\boldsymbol{\gamma}, \boldsymbol{\beta}} \rangle$ is tricky for the MIS problem even on bounded-degree graphs because the mixing Hamiltonian contains the exponential of noncommuting Pauli matrices. We therefore calculate numerically the expectation of the cost function $\langle C \rangle$ for the  square ring graph. Let us define $A= e^{-i \boldsymbol{\beta} H_M}e^{-i \boldsymbol{\gamma} H_C }$. For $p=1$  we have to calculate $\langle s  |A_1^\dagger C A_1| s \rangle$, where $| s \rangle$ is the initial state. We perform the numerical calculation for different choices of initial states. The expectation values for the independent sets (IS's) $|1000\rangle$, $|0100\rangle$, $|0010\rangle$, and $|0001\rangle$ are the same, and the expectation values for the maximal independent sets  $|0101\rangle$ and $|1010\rangle$ are the same. For the minimum depth QAOA circuit and $C_{max}=2$ we plot $\langle C_1 \rangle vs \beta_1$ as $\gamma_1$ cancels out of the expectation value. 

\begin{equation}
\langle C_1 \rangle = \frac{\langle s  |A_1^\dagger C A_1| s \rangle}{C_{max}}
\end{equation}

\begin{figure}[ht]
\centering
\includegraphics[width=3in , height =1.5 in]{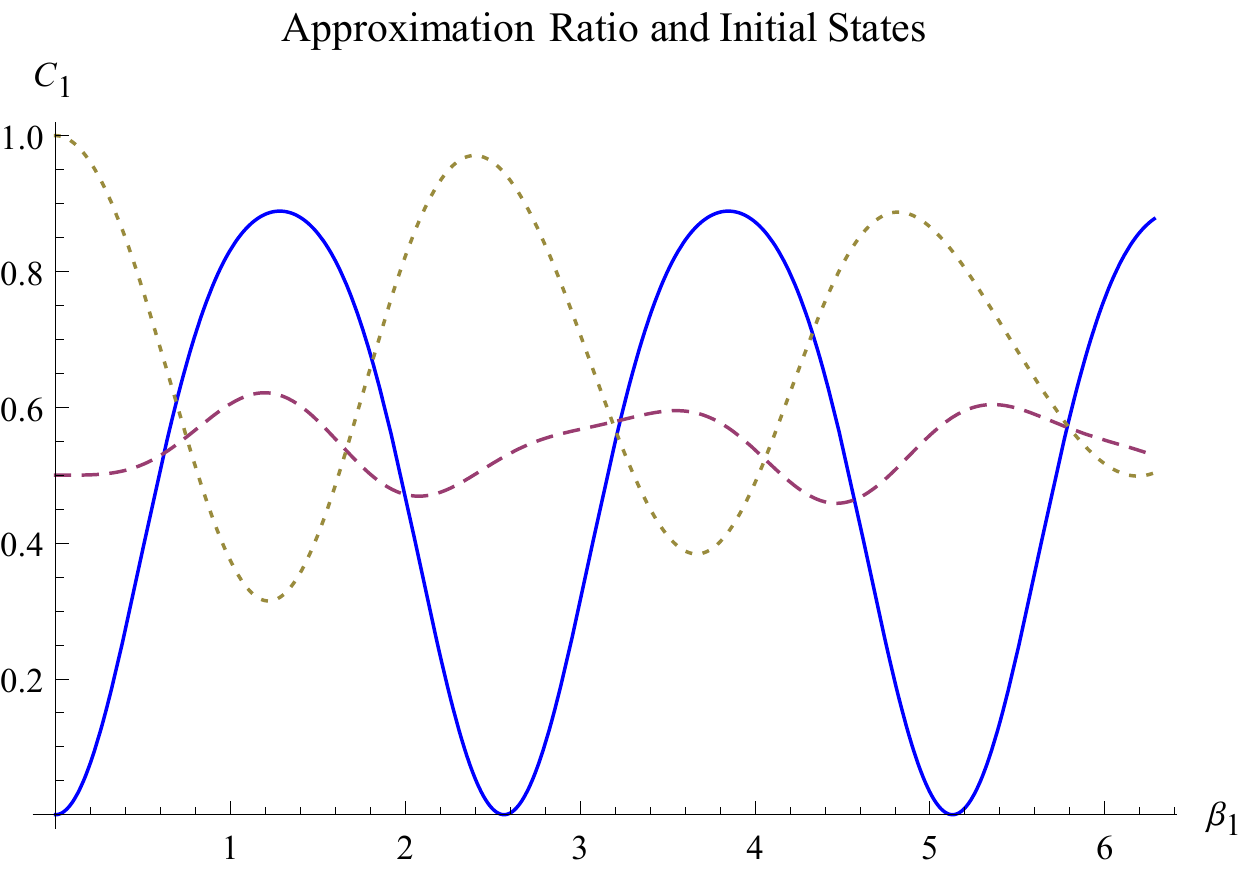}
\caption{approximation}
\end{figure}

The maximum value for the expectation $\max_{\gamma_1 , \beta_1} {\langle C_1 \rangle} = 1.0$, $0.89$ and $0.68$ for the MIS's, empty set, and IS's, respectively. We note that the approximation ratio is better for the empty set compared with the independent set states. 

\section{Conclusion}
We have studied the maximum weighted independent set problem using the quantum alternating operator ansatz. We note that the probability distribution of observance of the maximum independent states is asymmetric; in contrast, the probability distribution of the Max-Cut states is symmetrically distributed. We also calculated the approximation ratios for our graph for different initial states. In this paper we  considered a simple graph and observed the differences with the unconstrained problem. 

Much more research is needed in order to  understand our results analytically. We intend to run the experiments on larger graphs with larger circuit depths; and as the parameter space increases, it will be useful to understand improvements that can be made in the classical parameter optimization algorithms. We also plan to execute the algorithm on a quantum computer and see how far we can push it on a noisy intermediate-scale quantum device (NISQ).

Acknowledgments: I  thank Stuart Hadfield and Shengtao Wang for valuable discussions. This material was based upon work supported by the U.S. Department of Energy, Office of Science, under contract DE-AC02-06CH11357.

\end{document}